# Enhanced Backgate Tunability on Interfacial Carrier Concentration in Ionic Liquid-Gated MoS$_2$ Devices

*Qiao Chen[1], Chengyu Yan[*,1,2], Changshuai Lan[1], Qiyang Song[1], Yi Yan[1], Shun Wang[*,1,2]*

[1]MOE Key Laboratory of Fundamental Physical Quantities Measurement & Hubei Key Laboratory of Gravitation and Quantum Physics, National Gravimetry Laboratory and School of Physics, Huazhong University of Science and Technology, Wuhan 430074, China

[2]Institute for Quantum Science and Engineering, Huazhong University of Science and Technology, Wuhan 430074, China

Corresponding Authors: Chengyu Yan (chengyu_yan@hust.edu.cn); Shun Wang (shun@hust.edu.cn)

## Abstract

The periodic spatial modulation potential arising from the zig-zag distribution of ions at large gate voltage in an ionic liquid gated device may enable functionalities in a similar way as nanopatterning and moiré engineering. However, the inherent coupling between periodic modulation potential and carrier concentration in ionic liquid devices has hindered further exploration. Here, we demonstrate the feasibility of decoupling manipulation on periodic modulation potential and carrier density in an ionic liquid device by using a conventional backgate. The backgate is found to have a tunability on carrier concentration comparable to that of ionic gating, especially at large ionic liquid gate voltage, by activating the bulk channels mediated back tunneling between the trapped bands and interfacial channel.

## 1. Introduction

Imprinting a periodic spatial modulation potential, with a periodicity different from the lattice periodicity of the hosting material, to a two-dimensional electronic system has stimulated development in emerging optical and electronic functionalities[1-5] owing to the enhanced light-matter interactions[1, 6-8] and electron-electron correlations[5, 9, 10]. Nanopatterning and moiré engineering are the most common approaches to induce the periodic modulation[2, 11, 12]. Enormous efforts are being dedicated to these two approaches to enrich the tunability[13-17], including tuning the strength and periodicity of the modulation potential *in situ*. Interestingly, the *in situ* control over the strength and periodicity of the modulation potential can potentially be implemented via an alternative approach, i.e., ionic liquid gating at large gate voltage. Ionic liquid gating is most well-known for its capability of inducing massive amount of charge carriers at the interface[18-30], whereas its implication in generating periodic spatial modulation potential is somehow overlooked. Here, applying a large gate voltage will enforce the spatial distribution of ions undertake a transition from a uniform one to a zig-zag one, as illustrated in Figure 1a, to minimize the repulsion between the densely packed ions, which can be verified through topographic characterization[31-36], and theoretical simulation[37-42]. The zig-zag distributed ions in turn cast a spatially modulated potential to the two-dimensional system. The strength and periodicity of the modulation potential can be manipulated via the distribution of ions or equivalently speaking the gate voltage. Experimentally, the spatial modulation manifests itself via the formation of trapped bands or states at certain ionic liquid gate voltage[31, 43-46]. However, it is necessary to stress that the manipulation on the spatial modulation potential via ionic liquid gating does not warrant that one can realize similar functionalities enabled by nanopatterning or moiré engineering. In the case of nanopatterning or moiré engineering, the control over periodic modulation potential and carrier concentration, which are perhaps equally important in a wide range of applications, can be adjusted independently[13, 14]. In contrast, the manipulation on these two quantities is coupled together in the case of ionic liquid gating. It is therefore highly desirable to supplement the ionic liquid gating with additional control knobs.

Backgate seems to be a natural candidate at first glance since it provides extra control on the carrier concentration. However, from the perspective of capacitive effect, a conventional backgate has little impact in an ionic gated device, considering the electric field due to the backgate is negligible in comparison with the colossal one due to the ionic liquid gate[18, 30, 43, 47, 48]. Although there are also attempts of achieving sizable modulation on carrier concentration by unconventional backgates such as ionic solid gate[49, 50], the long-term stability and complementary metal oxide semiconductor (CMOS) compatibility of these unconventional backgates are still to be optimized. Fortunately, a conventional backgate can possibly participate the modulation of charge carriers in an indirect manner other than direct charging effect. For instance, it is demonstrated in a dual-gated polymer conductor that the weak electric field due to the backgate can perturb the Coulomb gap induced by ionic liquid, which results in a much larger change in the tunability compared to the expectation of the capacitive effect[44].

In this work, we demonstrate that a conventional backgate can lead to a modulation in the carrier concentration comparable to ionic liquid gate itself in a dual gated transition metal dichalcogenides (TMDs) device via indirect approach. It is indeed unlikely for the backgate to directly affect the interfacial channel, however, it can populate or deplete the bulk channels that mediate the back tunneling process between the interfacial channel and trapped bands, and hence notably tuning the transport property at the interface. As a result, our work highlights the possibility of decoupled control on periodic potential and carrier concentration in an ionic liquid gated device, and thus may accelerate advanced architecture that harnesses the extremely high carrier concentration and tunable periodic modulation potential at the same time.

## 2. Result and discussion

### 2.1 Emergence of Trapped Bands

$MoS_2$ flakes, etched into standard Hall bars, are dual-gated by an ionic liquid DEME-TFSI (N,N-diethyl-N-methyl-N-(2-methoxyethyl) ammonium bis

(trifluoromethylsulfonyl)-imide) on the top and a heavily doped Si backgate at the bottom. We first characterize the impact of ionic gate, with backgate grounded, in flakes with various thickness. Figure 1b depicts the transfer curve against the ionic gate voltage ($V_{ig}$) of a typical device with a thickness of 10 nm. It is seen that the drain current ($I_d$) initially rises with increasing $V_{ig}$ up to $V_{ig}$ = 3 V, then gradually decays and saturates, in line with previous reports[18, 51, 52]. The reduction and saturation of $I_d$ with large $V_{ig}$ are usually attributed to electrochemical reaction between the ionic liquid and devices under test[21, 44, 46]. However, it is ought to highlight that the electrochemical reaction is irreversible and thus the transfer curves should be diverse after repeating the measurements. Transfer curves in our experiments, on the other hand, almost overlay through different runs (see *Support Information* section 2 for details). Instead, the behavior in large $V_{ig}$ regime signifies the emergence of trapped bands, due to the spatial redistribution of ions driven by the ionic gate voltage, at the interface between the ionic liquid and MoS$_2$, analogous to the trapped bands observed in WS$_2$ devices[43]. This interpretation based on trapped bands can be verified experimentally.

First, localization of charge carrier becomes increasingly profound at larger $V_{ig}$. This effect will compete with the charge accumulation due to the conventional capacitive effect, and may therefore result in a non-monotonic evolution of the concentration $n_{2D}$ of mobile charge carrier with respect to $V_{ig}$. We indeed observed a non-monotonic $n_{2D}$-$V_{ig}$ relation, in good correlation with the evolution of $I_d$, as highlighted in Figure 1b. In addition, it is noteworthy that all transfer curve measurements in this study were systematically conducted at 220 K, whereas the $n_{2D}$ derived from Hall effect were obtained at 165 K (see *Experimental Section* for details).Second, the trapped bands can interplay with the interfacial states via an indirect back tunneling process mediated by the bulk channels. Here, the localized charge carriers of the trapped bands first hop into the partially populated bulk channels in the case of MoS$_2$, other materials may behave differently as we will demonstrate later on, and then return to the interfacial channels to contribute to $I_d$, as shown in Figure 1c. In this regard, the back tunneling effectively weakens the influence of localization. The probability of back tunneling process can be tuned by the number of bulk channels which is directly related to the thickness of the flakes in the case of MoS$_2$. It is then expected that $I_d$ at large $V_{ig}$ should be increased with enhanced back tunneling in a thicker flake with more bulk channels, since less charge carriers are effectively localized. This prediction is in good agreement with the experimental observation that $I_d/I_{peak}$, $I_{peak}$ is the maximum value of $I_d$ for a given flake thickness, saturates at larger value with increasing flake thickness, as shown by the raw data in Figure 1d or linear-rescaled data in Figure 1e after taken the variation in ionic liquid gate capacitance into account (see *Support Information* section 5 for details).

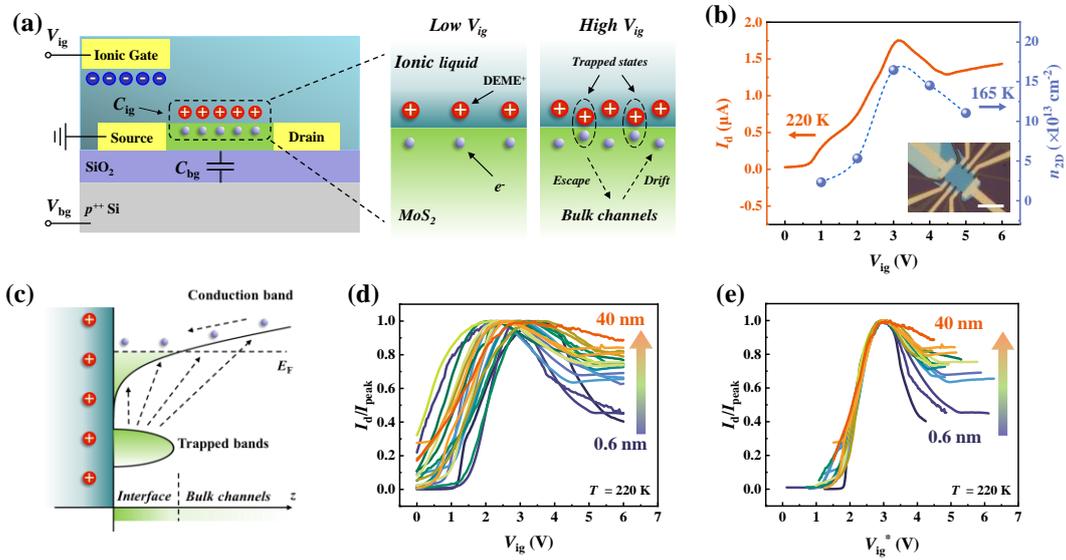

**Figure 1.** Trapped bands in an ionic-MoS$_2$ device and back tunneling process. a) Schematic representation of the ionic-MoS$_2$ device and ion distribution at the interface. The green channel denotes MoS$_2$ crystal, the blue spheres indicate anions TFSI$^-$, the red spheres represent cations DEME$^+$, and the grey spheres indicate electrons. Charge carriers are localized due to the zig-zag distribution of ions at large $V_{ig}$ as shown in the ellipses. b) Transfer characteristics curve ($V_d$ = 10 mV) and carrier density of the 10 nm ionic-MoS$_2$ device. Inset: the optical image of the device, the scale bar is 10 μm. c) Energy band of the ionic-MoS$_2$ device, with the *x*-axis representing the *z*-direction of the MoS$_2$ crystal. d) Transfer characteristic curves of ionic-MoS$_2$ devices with various thicknesses. e) Transfer characteristic curves after scaling $V_{ig}^* = α × V_{ig} − V_{offset}$ to eliminate the impact due to variation in the ionic liquid gate capacitance and intrinsic doping, here both $α$ and $V_{offset}$ are sample specific.

## 2.2 Tuning Back Tunneling Process with a Backgate

Interestingly, the back tunneling process opens a new venture to modulate the transport property of the interfacial channels with a backgate. The backgate can contribute via tuning the bulk channels. Apply a positive backgate voltage ($V_{bg}$) tends to populate the bulk channels and hence blocks the back tunneling, so that more charge carriers would be localized at the trapped bands and thus results in a reduction of $n_{2D}$ of mobile carriers. On the other hand, a negative $V_{bg}$ enhances the back tunneling process via depleting the bulk channels, which in turn leads to an increase in $n_{2D}$.

Our experimental results follow the prediction. The measurements with both ionic liquid gate and backgate activated were carried out at 165 K, far below the glass transition of ionic liquid DEME-TFSI. It is necessary to stress that backgate voltage is applied after the ionic liquid has already been frozen, hence the backgate voltage does not perturb the distribution of ions[53]. Let us first focus on the result in a flake with a thickness of 10 nm. At low $V_{ig}$, where the trapped bands are not well established, sweeping backgate voltage does not causes noticeable change in $n_{2D}$ evidenced as $\triangle n_{2D}$ is almost pinned to zero, $\triangle n_{2D}(V_{bg}, V_{ig}) = n_{2D}(V_{bg}, V_{ig}) - n_{2D}(0, V_{ig})$, regardless of the polarity of $V_{bg}$ in Figure 2a. A weak linear dependence of $\triangle n_{2D}$ on $V_{bg}$ is discernible upon closer examination, originating from the conventional capacitive coupling of the backgate. Gradually increase $V_{ig}$ will drive the formation of trapped bands. With the formation of trapped bands, it is found that $V_{bg}$ can change $\triangle n_{2D}$ dramatically. Additional carrier concentration on the order of $10 \times 10^{13}$ cm$^{-2}$ can be supplemented to the system at $V_{ig} > 3$ V (when trapped bands is firmly established) with negative $V_{bg}$, comparable to the carrier concentration induced by the ionic liquid gate itself at the corresponding voltage. Positive $V_{bg}$ leads to a reduction of carrier concentration on the order of $1 \times 10^{13}$ cm$^{-2}$ at large $V_{ig}$. The observation that negative $V_{bg}$ adds charge carriers whereas positive one removes carriers agrees with the back tunneling model, but in staggering contrast to the standard functionality of a backgate based on the capacitive model. The aforementioned trend may be more easily recognized after presenting the data in a complementary manner, i.e., $n_{2D}$ as a function of $V_{ig}$ at fixed $V_{bg}$, as depicted in Figure 2b. It is clear that traces with positive $V_{bg}$ mimic the intrinsic behavior of an ionic liquid tuned device ($V_{bg} = 0$ V) until $V_{ig} \sim 3$ V, then reduction in $n_{2D}$ due to trapped bands becomes more effective with positive $V_{bg}$. On the contrary, negative $V_{bg}$ not only induces sizable additional charge carrier, pushing $n_{2D}$ to an enormous value of $35 \times 10^{13}$

cm$^{-2}$, but also delays the full dominance of the trapped bands to $V_{ig}$ = 4 V.

Qualitatively similar results have been systemically verified in flakes with thickness spanning from 1.8 nm to 80 nm as summarized in Figure 2c & d. The effectiveness of the backgate is rather sensitive to the thickness of the flakes. The backgate is most efficient in flakes with an intermediate thickness around 10 nm, whereas its impact decays when the thickness approaches both few-layer limit and bulk limit. In the few-layer limit, there is rather limited number of bulk channels to start with, the back tunneling of the localized charge carriers is quite restricted regardless of the occupation of the bulk channels, hence the backgate is not influential. In the bulk limit, the backgate primarily modules the bulk channels near the bottom surface, its tunability on the bulk channels in the close vicinity of the ionic liquid-MoS$_2$ interface, that really matters to the back tunneling process, is largely screened, hence the backgate is again non-efficient. Nevertheless, the backgate tunability still exceeds that expected from the capacitive effect even in the few-layer limit or bulk limit, companied with reversed polarity.

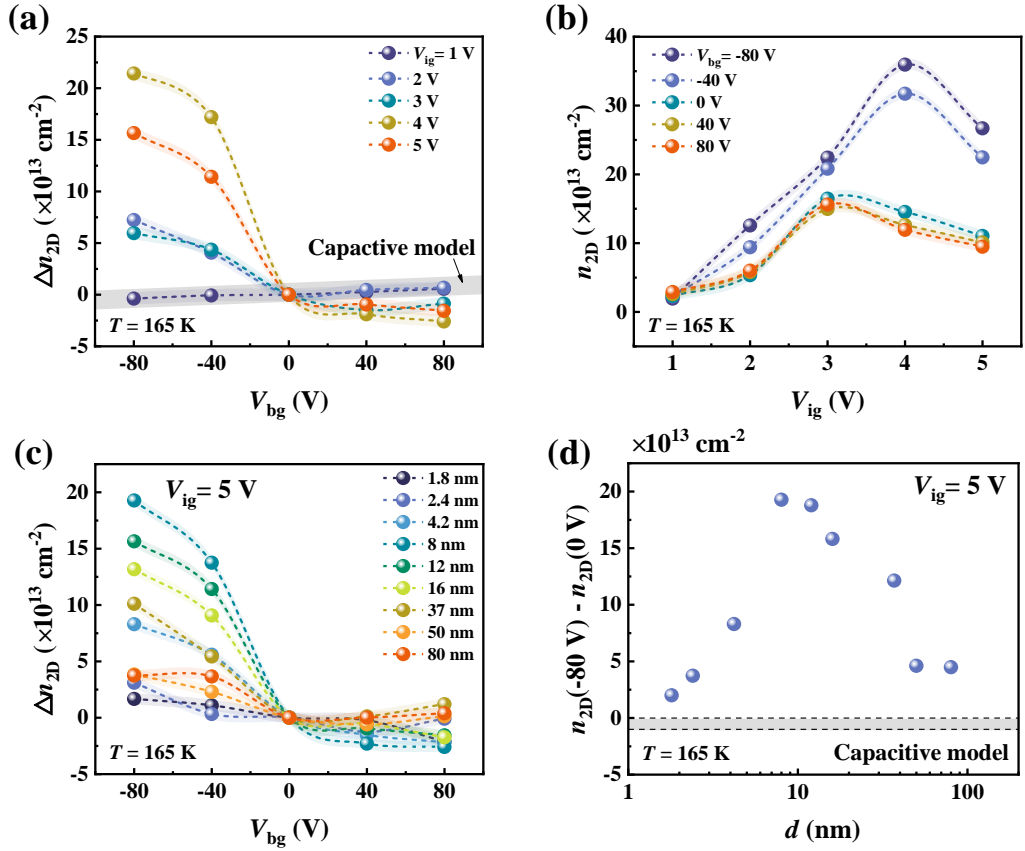

**Figure 2.** Demonstration of efficient tunability on carrier concetration via a backgate at 165 K. a) The response of relative carrier concetration $\triangle n_{2D}$ to $V_{bg}$ with incremented $V_{ig}$, $\triangle n_{2D}(V_{bg}, V_{ig}) = n_{2D}(V_{bg}, V_{ig}) - n_{2D}(0, V_{ig})$, in a 10 nm flake, grey strip indicates the tunability based on the conventional capactive model of bakcgate. b) The response of $n_{2D}$ to $V_{ig}$ with fixed $V_{bg}$. c) $\triangle n_{2D}$ as a function of $V_{bg}$ at $V_{ig}$ = 5 V in flakes with different thickness. d) Backgate tunability as a function of flakes thickness at $V_{bg}$ = -80 V, grey strip indicates the tunability based on the conventional capacitive model of bakcgate.

## 2.3 Necessary Electronic Properties of Achieving Notable Backgate Tunability

Since the presence of the bulk channels and their occupancy are crucial to the back tunneling process, it is interesting to analyse the behavior in different TMDs especially in large $V_{ig}$ regime. Here, we take $WS_2$ and $WSe_2$ as examples. Trapped bands also forms in these two materials at large $V_{ig}$ in a similar manner as that observed in $MoS_2$ (see *Support Information* section 7 for details). However, thickness dependence data and backgate tunability data shows qualitatively different behavior in $WS_2$ and $WSe_2$ with respect to $MoS_2$.

Figure 3 summarizes the thickness dependence data. Figures 3a & b show the transfer characteristics of WS$_2$ and WSe$_2$ devices using the same linear-rescaling as for MoS$_2$ (see *Support Information* Section 5 for details). Unlike MoS$_2$, the saturate $I_d$ in WS$_2$ and WSe$_2$ devices is insensitive to thickness, whereas MoS$_2$ shows a logarithmic-like thickness dependence (Figure 3c). It is also interesting to note that the normalized saturation current in WSe$_2$ flakes is noticebly smaller than that in MoS$_2$ or WS$_2$ flakes with similar thinkness.

Figure 4 encloses the backgate tunability data in WS$_2$ and WSe$_2$. In the presence of negative $V_{bg}$, the carrier contentration decreases with decreasing gate voltage in both WS$_2$ and WSe$_2$ following the expectation of capacitive model, which is opposite to the trend in MoS$_2$. Besides, the change in the carrier concentration is on the order of $1 \times 10^{13}$ cm$^{-2}$ in WS$_2$ and WSe$_2$, an order smaller than that in MoS$_2$. On the other hand, similar behavior has been observed in all the three materials in the presence of positive $V_{bg}$, apart from change in carrier concentration in MoS$_2$ and WS$_2$ are comparable and about 2 times larger than that in WSe$_2$.

The difference behavior in WS$_2$ and WSe$_2$ in comparison with MoS$_2$ can be explained qualitatively by the back tunneling process. First, WSe$_2$ has a much larger interlayer resistance than MoS$_2$[54-56]. This is perhaps also the case for WS$_2$, since higher carrier concentration dramatically reduces interlayer resistivity[56, 57]. When comparing MoS$_2$ with intrinsic doping induced by sulfur vacancy to WS$_2$ and WSe$_2$ which remain depleting intrinsically, it is natural that MoS$_2$ holds the smallest interlayer resistivity than WS$_2$ and WSe$_2$. Furthermore, characterization of interlayer resistivity through top-electrode/TMDs/bottom-electrode devices provides additional validation of the discussion above (see *Support Information* Section 8 for details). The back tunneling process in WS$_2$ and WSe$_2$ primarily occurs through the bulk channel in the vincinity of the interface whereas other bulk channels are well isolated from the interface, hence the saturation current is almost thickness independent. Second, the bulk channels in WS$_2$ and WSe$_2$ are almost empty in the absence of backgate voltage whereas they are partially occuped in MoS$_2$. Hence, a negative $V_{bg}$, which tries to deplete the chaneles, has little impact on the already almost empty bulk channels in WS$_2$ and WSe$_2$ in terms of the back tunneling process, instead, the trivial capacitive effect becomes explicit.

The bulk channels are far from fully occupied with the largest backgate voltage in our exerpiemnt, hence positive $V_{bg}$ leads to similar results in all the three materials.

Through the comparison of these three TMDs, it hightligths that enhance backgate tunability may be activated in material with suitable interlayer resistance and intrinsically partially occupied bulk channels. A more quantitative analysis would be benficial to further optimise the backgate tunability, however, it is likely to require additional inputs such as the binding energy between the ions and the charge carrier in addition to the aforementioned interlayer resistance and instrinsic bulk channels occupancy, which calls for further investigation.

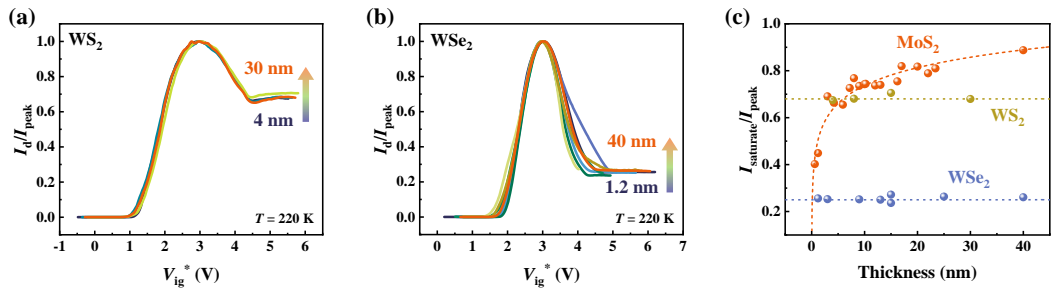

**Figure 3.** Thickness dependence of back tunneling process in $WS_2$ and $WSe_2$. a) & b) Rescaled transfer characteristic curves of $WS_2$ and $WSe_2$ device with various thicknesses, respectively. c) Comparison of the normalized saturation current in the three materials. Normalized saturation current is insensitive to flake thickness in $WS_2$ and $WSe_2$, whereas it shows a logarithmic-like reduction as a function of thickness in $MoS_2$.

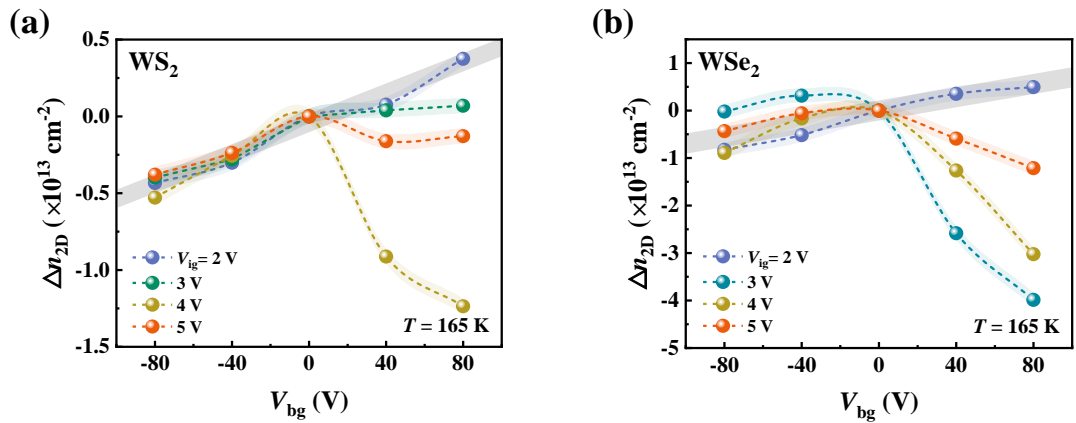

**Figure 4.** Backgate tunability in WS$_2$ and WSe$_2$ with a flake thickness of 10 nm. a) & b) $\triangle n_{2D}$ as a function of backgate voltage with incremented $V_{ig}$ in WS$_2$ and WSe$_2$, respectively, grey strip indicates the tunability based on the conventional capacitive model of bakcgate.

## 3. Conclusion

This work demonstrates the enhanced backgate tunability on interfacial carrier concentration in an ionic liquid-gated MoS$_2$ device by a conventional backgate. The main role of the backgate gate is tuning the bulk channels that mediates the back tunneling between the trapped bands forming at large ionic liquid gate voltage and interfacial channels, instead of inducing charge carrier via the trivial capacitive effect. The enhanced tunability of the backgate is key in utilizing the naturally existed periodic spatial modulation in an ionic liquid gated device to full potential, by addressing the issues of lacking of independent control on carrier concentration and periodic modulation potential.

## 4. Experimental Section

*Crystal Exfoliation and Characterization*: MoS$_2$ films were obtained from bulk MoS$_2$ (HQ Graphene Inc.) by mechanical exfoliation using Scotch tape and transferred onto 285 nm SiO$_2$/*p*-doped Si substrates[58]. The thickness of the MoS$_2$ films was calibrated by atomic force microscope (AFM) (Bruker Dimension Edge Inc.), Raman spectroscopy, and optical microscope[59].

*Device Fabrication*: MoS$_2$ flakes were selected and fabricated using a two-step EBL process, which utilized a bilayer resist consisting of Copolymer EL6 and PMMA 495A8. Following the first electron beam lithography (EBL, Crestec Inc.) step, standard Hall bar shapes were etched using reactive ion etching (RIE, Oxford Inc.)[60]. After the second EBL step, Ti/Au (5/60 nm) was deposited and subsequent lift-off using acetone.

*Electrical Measurement*: All electrical measurements were conducted in a cryostat (TeslatronPT, Oxford Inc.). Transfer curve measurement utilized the Keithley 2634B source-measure unit, while the Hall effect and other transport experiments employed the Keithley 6221 current source and 2182A nanovoltmeter. It is necessary to stress that we have etched the flakes in to the shape of standard Hall bar to accurately quantify the carrier concentration. All transfer curve measurements were systematically conducted at 220 K, slightly above the glass transition temperature (~190 K) of ionic liquid DEME-TFSI, to suppress interfacial electrochemical reactions[21, 23], since the electrochemical window of ionic liquid exhibits significant temperature dependence, expanding from 3 V at 300 K to over 6.5 V at 220 K for DEME-TFSI[51]. Concurrently, all the $n_{2D}$ were acquired by Hall effect measurements at 165 K under continuous $V_{ig}$ application throughout the cooling protocol (from 220 K to 165 K). This cryogenic protocol ensures frozen ion configurations, thereby eliminating dual-gate coupling effects[53, 61] and enabling accurate determination of $n_{2D}$-$V_{bg}$ relationships.

# Acknowledgement

We acknowledge the financial support from the National Natural Science Foundation of China (Grant No. 12204184, 12074134).

# Keywords